\documentclass[prd,aps,a4paper,nofootinbib,twocolumn]{revtex4}

\newif\ifusesec
\usesectrue  
   
\usepackage{graphicx} 
\usepackage{mathrsfs}
\usepackage{latexsym,amsmath,amsfonts,amssymb}
\usepackage{multirow}

\newcommand{\be}{\begin{equation}}
\newcommand{\ee}{\end{equation}}
\newcommand{\bea}{\begin{eqnarray}}
\newcommand{\eea}{\end{eqnarray}}

\newcommand{\D}{\partial}
\newcommand{\cL}{\mathcal{L}}
\newcommand{\La}{\Lambda}
\newcommand{\G}{\Gamma}
\newcommand{\vS}{{\bf S}}
\newcommand{\Tf}{T_{\rm eff}}
\newcommand{\Ef}{E_{\rm eff}}
\newcommand{\pcm}{{p_{\rm c.m.}}}

\def\dfrac{\frac}
\def\p{{\mathbf p}}
\def\n{{\mathbf n}}

\begin{document}

\title{Gravitational spin-orbit coupling in binary systems, \\ post-Minkowskian approximation and effective one-body theory}

\author{Donato \surname{Bini}$^1$}
\author{Thibault \surname{Damour}$^2$}
 
\affiliation{$^1$Istituto per le Applicazioni del Calcolo ``M. Picone'', CNR, I-00185 Rome, Italy\\
$^2$Institut des Hautes Etudes Scientifiques, 91440 Bures-sur-Yvette, France}

\date{\today}

\begin{abstract}
A novel approach for extracting gauge-invariant information about spin-orbit coupling in gravitationally interacting binary systems
is introduced. This approach is based on the ``scattering holonomy", i.e. the integration (from the infinite past to the infinite future)
of the differential spin evolution along the two worldlines of a binary system in hyperboliclike motion. 
We apply this approach to the computation, at the first post-Minkowskian
approximation (i.e. first order in $G$ and all orders in $v/c$), of the  values of the two gyrogravitomagnetic ratios describing 
spin-orbit coupling in the Effective One-Body formalism. These gyrogravitomagnetic ratios are found to tend to zero in the
ultrarelativistic limit.
\end{abstract}

\maketitle

\section{Introduction} \label{sec1}

The Effective One-Body (EOB) formalism was conceived \cite{Buonanno:1998gg,Buonanno:2000ef,Damour:2000we,Damour:2001tu}
with the aim of analytically describing both the last few orbits of, and the complete gravitational-wave signal emitted by, coalescing
binary black holes. The EOB formalism played a key role in allowing one to compute, in a semi-analytic way,  hundreds of thousands
of templates which have been used to search for, and analyze, the recently detected gravitational wave signals from
coalescing binary black holes \cite{Abbott:2016blz,Abbott:2016nmj,Abbott:2017vtc}. Any theoretical advance in EOB theory might benefit to the burgeoning field of gravitational wave astronomy.
The present work will introduce a new approach to the theoretical description of spin-orbit couplings within the EOB formalism,
and apply it to the computation, to the first post-Minkowskian (1PM) approximation (i.e. first order in $G$ but {\it all orders in} $v/c$), 
of the  values of the two gyrogravitomagnetic ratios, $g_S$ and $g_{S_*}$, which describe spin-orbit coupling in EOB theory.

The EOB formalism was originally developed within the post-Newtonian (PN) approximation method \cite{Buonanno:1998gg,Buonanno:2000ef,Damour:2000we,Damour:2001tu}.
Recently, a novel approach  to EOB theory, based on the post-Minkowskian (PM) approximation method, has been introduced \cite{Damour:2016gwp}. The PM approximation scheme (expansion in powers of $G$, for fixed velocities) physically correspond to small-angle
scattering in hyperboliclike encounters (with arbitrary incoming velocities) 
\cite{Westpfahl:1979gu,Portilla:1979xx,Portilla:1980uz,Bel:1981be,Damour:1981bh,Westpfahl:1985,Westpfahl:1987,Ledvinka:2008tk}.
Ref. \cite{Damour:2016gwp} has applied this approach  to the orbital dynamics of a system of two nonspinning
compact bodies. The aim of the present paper is to generalize this approach to the case of gravitationally interacting {\it spinning} bodies, considered in the pole-dipole approximation, i.e. described by two masses ($m_1, m_2$) and two spin 4-vectors ($s_1, s_2$).
We will work linearly in the spins, and consider (as in \cite{Damour:2016gwp}) hyperboliclike motions. 

When considering nonspinning binary systems, the prime  observable
consequence of the orbital dynamics is the  scattering angle $\chi$, measured in the center-of-mass (c.m.).
More precisely, as emphasized in Refs. \cite{Damour:2009sm,Damour:2014afa}, the orbital dynamics
is encoded, in a {\it gauge-invariant} manner, in the functional link between $\chi$ and the total (c.m.) energy, 
${\mathcal E}_{\rm real}$, and (c.m.) orbital angular momentum, $L$, of the system. When considering {\it spinning} systems,
with parallel spins, one has to deal with a more general (gauge-invariant) functional link, namely
$\chi= \chi(L, S_1, S_2)$, where $S_1, S_2$ denote the algebraic magnitudes of the parallel spins \cite{Bini:2017wfr}.
[The orbital angular momentum, and the spin magnitudes are defined so that $J= L+S_1+S_2$ is equal to
the (well-defined)  total angular momentum of the system in the c.m. frame.] See Ref. \cite{Bini:2017wfr} for
 recent high PN-order results on the function $\chi(L, S_1, S_2)$.

Here we introduce an alternative approach to a gauge-invariant characterization of the dynamics of
spinning systems.  Instead of being based on a scalar function [$\chi(L, S_1, S_2)$] that is only
defined for parallel spins, we will instead consider {\it matrix-valued} (gauge-invariant) observables:
the ``scattering holonomy," and the related ``spin holonomy" (both being defined along each of the
two infinite worldlines representing the spacetime history of the two bodies). These quantities will be
defined in the following Sections. Before entering any technical detail, let us emphasize that a significant
advantage (over the computation of $\chi(L, S_1, S_2)$) of our new method, is that we will get information about the
linear-in-spin couplings from a calculation where we will be able to actually neglect spin-effects in the dynamics !
This useful feature of our new approach is akin to the method used in Refs. \cite{Damour:2007nc,Damour:2008qf}
to derive (within a PN framework) the spin-orbit terms in the two-body Hamiltonian from the metric generated
by two nonspinning bodies.

In the present paper, we apply our approach to the computation of the spin-orbit couplings at the
linear order in $G$ (1PM approximation). In addition, we show how to extend the dictionary between the
real two-body dynamics and its EOB image so as to allow us to transcribe our explicit 1PM spin-scattering
computation, into a corresponding 1PM-accurate knowledge of the two gyrogravitomagnetic ratios, $g_S$ and $g_{S_*}$
that describe spin-orbit coupling in EOB theory. 

We (generally) employ units where $c=1$; use the mostly plus
signature for the spacetime metric; and  use standard EOB notations, notably
\be
M \equiv m_1 +m_2 \,; \, \mu \equiv \frac{m_1 m_2}{m_1+m_2} \, ,
\ee 
 with the symmetric mass ratio of the binary system being denoted as
 \be
 \nu  \equiv \frac{\mu}{M} = \frac{m_1 m_2}{(m_1+m_2)^2} \, .
 \ee

\section{New concept: scattering holonomy} \label{sec2}

We consider the scattering of two, gravitationally interacting, spinning bodies. 
Geometrically, we deal with two worldlines, $\cL_1$ and $\cL_2$, in an asymptotically flat curved spacetime endowed
with a metric $g=g_{\mu \nu}dx^\mu \, dx^\nu$ (generated by the energy tensor supported on the worldlines). 
$\cL_1$ and $\cL_2$ asymptote to straight lines
in the infinite past and the infinite future. In our pole-dipole approximation, each worldline (say, $\cL_1$) is {\it a priori} endowed 
with three 4-vectors, namely : the 4-velocity, $u_1=u_1^\mu \D_\mu$, 
(with the usual unit normalization $u_1 \cdot u_1=-1$), the 4-momentum $p_1=p_1^\mu \D_\mu$ (with the normalization
$p_1 \cdot p_1=-m_1^2$) , and a spin 4-vector, $s_1=s_1^\mu \D_\mu$ (constrained to satisfy $p_1 \cdot s_1=0$,
and $s_1 \cdot s_1 = s_1^2=$ cst.).
[Such a 4-vector $s_1$ is equivalent to an antisymmetric spin tensor $S_1^{\mu \nu}$ satisfying $S_1^{\mu \nu} p_{\nu}=0$.
In a local Lorentz frame\footnote{In a general frame $S_1^{\mu \nu} = m_1^{-1}\eta^{\mu \nu \kappa \lambda} p_{1 \kappa} s_{1 \lambda}$, with the appropriate definition of the Levi-Civita tensor $\eta$.} whose unit time vector is $p_1/m_1$, the (only nonvanishing) spatial components of $S_1$ are 
defined to be dual to the (only nonvanishing) spatial components of $s_1$: $S_1^{\hat i \hat j}= \epsilon^{\hat i \hat j \hat k} s_1^{\hat k}$.]
The classic works (among others) of Mathisson, Papapetrou, Tulczyjew and Dixon (see \cite{Dixon:1970zza}
and references therein) have shown that the data $u_1, p_1, s_1$ (or $u_1, p_1, S_1$) satisfy a set of  universal evolution equations.
The latter equations imply, in particular, a link between $u_1$ and $p_1$ of the form
\be
p_1= m_1 u_1 + O(s_1^2)\,.
\ee 
Using this link, the differential evolution system for $p_1$ and $s_1$ has the form 
\be
\frac{D_g p_{1 \mu}}{d \tau_1}= -\frac12 R_{\mu \nu \kappa \lambda} u_1^\nu S_1^{\kappa \lambda}\,,
\ee
and
\be
 \frac{D_g s_1^\mu}{d \tau_1}= O(s_1^2)\,.
\ee
Here, $D_g$ denotes the covariant derivative (along the worldline) associated with $g=g_{\mu \nu}dx^\mu \, dx^\nu$, and $R_{\mu \nu \kappa \lambda}$
the corresponding curvature tensor (with the usual sign convention that $R_{1212}>0$ on a 2-sphere).

Because we work linearly in spins, and because
the method we shall introduce extracts the spin-orbit coupling
 only from the evolution of $s_1$ along $\cL_1$, we are allowed (as will be clear from our formulas below) to neglect 
 the $O(s_1)$ spin-curvature coupling in the evolution of $p_1= m_1 u_1 + O(s_1^2)$, and use a simplified evolution system
amounting to stating that  $p_1= m_1 u_1$, and that $u_1$ and $s_1$ are parallely propagated (in $g_{\mu \nu}$) along $\cL_1$ 
(with  a corresponding statement for $u_2$ and $s_2$ along $\cL_2$). Using an index-free notation, and working with differential forms,
we will write
\be \label{evolutionsystem}
D_g u_1=0 = D_g s_1 \,,
\ee
where
\be
D_g = d + \omega_1.
\ee
Here, the differential $d$ is taken along  $\cL_1$, and $\omega_1$ denotes the evaluation along $\cL_1$ of the
Levi-Civita connection one-form $\omega$ acting on contravariant four-vectors. 
Henceforth, we generally think of $u_a$, $s_a$ ($a=1,2$), $g$, $\omega$ as abstract geometric objects. 
These objects can then be expressed in terms
of various types of components: e.g. coordinate-frame components when using a generic coordinate system, or moving-frame components
when using such frames. In a coordinate frame the connection, acting on a contravariant four-vector (e.g. $u_1=u_1^\mu \D_\mu$, 
or $s_1=s_1^\mu \D_\mu$) becomes
\be
\omega^\mu_{\; \; \nu}=\G^\mu_{\; \; \nu \lambda}\, dx^\lambda\,,
\ee
where
\be
\G^\mu_{\; \; \nu \lambda}= \frac12 g^{\mu \sigma} \left( \D_\nu g_{\lambda \sigma} + \D_\lambda g_{\nu \sigma} - \D_\sigma g_{\nu \lambda} \right)\,.
\ee

To close the evolution system \eqref{evolutionsystem}, we need to replace $g_{\mu \nu}$ by the solution of
Einstein's equations, with the corresponding pole-dipole source terms supported by the two worldlines.
We also need, as is standard in perturbative approaches to the two-body problem, to regularize the formally infinite
evaluations of $g$ and $\omega$ along the worldlines. [See Ref. \cite{Bel:1981be} for a detailed discussion of
these regularizations within the PM context.] Let us note in advance another simplifying feature of our approach.
As  we can consistently neglect
the Mathisson-Papapetrou, $O(s_1)$, non-geodesic correction to $D_g u_1/d \tau_1$, we can also (when 
considering the generation of $g_{\mu \nu}$ by our spinning binary system) neglect the spin contributions to $g_{\mu \nu}$,
i.e. consider that $g_{\mu \nu}$ is generated by two nonspinning point masses.

We now define the {\it scattering holonomy}, $\La_1$, along $\cL_1$ as the parallel-transport linear operator (acting on contravariant four-vectors)
integrated along $\cL_1$ from the infinite past to the infinite future. As a parallely transported vector $v$ satisfies
\be
d v= - \omega_1 \, v
\ee
(where $\omega_1$ is evaluated on $\cL_1$) the $\cL_1$ scattering holonomy reads ($T$ denoting Dyson's time-ordered product \cite{Dyson:1949bp})
\bea \label{La11}
\La_1&=& T_{\cL_1} \left[ e^{-\int \omega_1} \right]\nonumber \\
&=& 1 - \int_{-\infty}^{+\infty}\omega_1 + \frac12 \int_{-\infty}^{+\infty} \int_{-\infty}^{+\infty} T \left[ \omega_1 \omega'_1 \right] + \cdots
\eea
with a similar result along $\cL_2$ given by exchanging $1 \leftrightarrow 2$. The time-ordered product in Eq. \eqref{La11}
refers to an integration performed along $\cL_1$, from the infinite past to the infinite future. In Eq. \eqref{La11},
$ \omega_1= \omega_1(t) = \sigma_1(t) dt$, $ \omega'_1= \omega_1(t') = \sigma_1(t') dt'$, and
$T \left[ \omega_1 \omega'_1 \right] = \sigma_1(t_>) \sigma_1(t_<) dt dt'$, with $t_<= {\rm min}(t,t')$
and   $t_>= {\rm max}(t,t')$ denotes the time-ordered product of two integrands coming from the formal expansion of
the exponential in the first line. Here, $t$ denotes any time-related parameter along $\cL_1$. [Note that our differential-form
formulation does not depend on the choice of any specific parametrization along the worldline.]

The {\it scattering holonomy}, $\La_1$, is a linear operator mapping the (abstract) vector space of contravariant four-vectors
at $-\infty$ (along $\cL_1$) onto its analog vector space at $+\infty$. As we describe here the scattering motions of an isolated system, 
the two latter asymptotic spaces of four-vectors can be naturally identified with the vector space of Minkowski four-vectors. [We use here
asymptotic flatness.] In other words, if we use a coordinate system that respects manifest asymptotically flatness (as will be the case in our PM computation),
the scattering holonomy computed as in Eq. \eqref{La11} will concretely be a $4 \times 4$ matrix  ${\La_1}^\mu_{\, \,\nu}$ acting
on Minkowski vectors. 

It will be henceforth convenient to denote by a subscript $-$ (respectively $+$) asymptotic quantities at $-\infty$  (resp. $+\infty$).
The {\it scattering holonomy} is a linear map between $u_1^-, s_1^-$ and $u_1^+, s_1^+$:
\be \label{scatteringoperator}
u_1^+= \La_1 u_1^- \, ;\, s_1^+= \La_1 s_1^-  \,.
\ee
This linear operator is geometrically defined and therefore, because of asymptotic flatness,  gauge-invariantly defined. [Concretely
the matrix ${\La_1}^\mu_{\, \,\nu}$ is invariant under coordinate diffeomorphisms that decay sufficiently fast at infinity 
so as to respect manifest asymptotic flatness.]
$ \La_1$ is a classical scattering operator which describes the mapping between the incoming momentum\footnote{The curvature-related
$O(s_1^2)$ general difference between $p_1$ and $m_1 u_1$ tends asymptotically to zero.}, $p_1^-=m_1 u_1^- $
and spin $s_1^-$, and the corresponding outgoing ones, $p_1^+=m_1 u_1^+ $
and $s_1^+$. Note that it contains in particular the information about the usual scattering angle $\chi$. As we have
neglected the spin corrections to the evolution of $u_1$, our present estimate of $ \La_1 $ only describes (when acting on $u_1^-$)
the orbital part, $\chi_{\rm orb}(L)$, of the scattering angle. However, we shall see how to extract spin-orbit information from the
action of $ \La_1 $ on spacelike vectors.

As we said, all the asymptotic four-vectors $u_a^{\pm}, s_a^{\pm}$ ($a=1,2)$ live in an asymptotic Minkowski space. 
Moreover, as parallel transport preserves the length, the two $\La_a$'s preserve the asymptotic flat metric. In other words,
the two matrices ${\La}^{\; \mu}_{a\; \; \nu}$  are
usual Lorentz transformations belonging to $SO(3,1)$, and preserving $\eta_{\mu \nu} = {\rm diag}(-1,+1,+1,+1)$.

\section{``Spatial" spin vectors}

The covariant spin four-vectors $s_1$, $s_2$ are not the usual,  canonical, spatial spin three-vectors $\vS_1, \vS_2$ (with
constant Euclidean lengths) that enter the Hamiltonian
description of binary dynamics (such as the EOB description). A simple way of constructing spatial spin three-vectors associated with
the four-vectors $s_a$ (which are orthogonal to the tangent $u_a$ to $\cL_a$)\footnote{At linear order in spins we do not need to
distinguish between orthogonality to $u_a$ or to $p_a$ because $p_a=m_a u_a +O(s_a^2)$.} has been given in Ref. \cite{Damour:2007nc}.
It can be described geometrically in the following way. Given two (future-directed) unit timelike vectors $u$ and $v$ in the tangent space
of some point in a Riemannian manifold, there is a unique local Lorentz transformation, say $B_g(u \!\!\to \! v)$ (where the letter $B$
stands for ``boost"), which acts in the 2-plane spanned by $u$ and $v$ 
so as to rotate $u$ into $v$, while leaving invariant the complementary 2-plane orthogonal to $u$ and $v$. [We give below the
explicit expression of the $4 \times 4$ boost matrix $\left[B_g(u \to v)\right]^\mu{}_\nu$ in the case where the metric $g$ is flat.]
With this notation in hand, and given a global field of  (future-directed) unit time-vectors $U = U^\mu \D_\mu$ orthogonal
to the time-slicing that we use to describe our spacetime\footnote{One could relax this conceptually simplifying condition.} , 
the spatial spin vector  associated with the four-vector $s_1$ is defined, at each point $x_1 \in \cL_1$, by (actively) applying the boost operator
$B_g(u_1 \!\!\to \! U)$,  
acting in the tangent vector space at the point $x_1$.
Note that $B_g(u_1 \!\!\to \! U)$ is defined 
so that it rotates $u_1$ into $U$, i.e. it is the inverse of
the boost that would map the lab frame associated with $U$ into the local rest-frame of $\cL_1$. We have indicated in subscript that
this linear map is locally defined in a curved spacetime with metric $g$ (evaluated at point $x_1$).

We then define\footnote{To ease the notation, we do not explicitly indicate that the vector $U$
entering the boost operator denotes the local value $U(x_1)$ of the global field $U(x)$.} 
\be
S_1 = B_g(u_1 \!\!\to \! U) \, s_1 \, ; \, S_2 =  B_g(u_2 \!\!\to \! U) \, s_2 \,,
\ee
where the abstract four-vectors $S_1$, $S_2$ now live in the corresponding  local three-planes orthogonal to $U(x_1)$, $U(x_2)$.
See Fig. 1, which assumes that the vector field $U$ is globally orthogonal to a spacelike hypersurface,
as is the case in the Hamiltonian formalism where spacetime is foliated (in the c.m. frame) by $t=$cst. 
hypersurfaces. As discussed next, only the flat-spacetime asymptotic limit (at $\pm \infty$) of $U$ 
will enter our results.

\begin{figure}
\includegraphics[scale=0.30,angle=90]{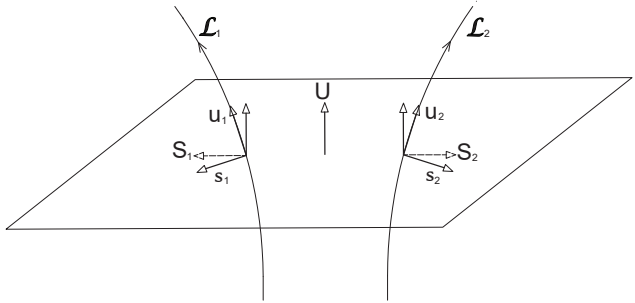}
\caption{\label{fig:1} Schematic representation of the definition of the  (abstract) spatial spin vectors $S_a$ ($a=1,2)$ orthogonal 
to the time direction $U$ defining the c.m. frame. They are obtained from the covariant spin vectors $s_a$ (orthogonal to 
$p_a = m_a u_a + O(s^2)$) by the Lorentz boost that rotates $u_a$ into the local value of $U$ (drawn vertically), while leaving
invariant the 2-plane orthogonal to $u_a$ and $U$. 
For ease of drawing,
we have sketched the scattering geometry as if the interaction were repulsive rather than attractive, and we have
represented orthogonality relations as if we were in an Euclidean space.}
\end{figure}

When decomposed  with respect to (wrt)
some suitably Cartesianlike orthonormal frame, say $U, E_1, E_2, E_3$, we have $S_a = S_a^1 E_1+  S_a^2 E_2+ S_a^3 E_3$,
where the three components $S_a^1, S_a^2, S_a^3$ define the three vector $\vS_a$. By construction the Euclidean length of 
$\vS_a$ is equal to the (constant) $g$-measured length of $s_a$, and is therefore also constant. 

When discussing the scattering operator \eqref{scatteringoperator}, only the asymptotic values (at $\pm \infty$) of the 
spin four-vector $s_1$ matter. The corresponding asymptotic values of the spatial spin vector $S_1$  are given by
\be \label{S1as}
S_1^+ = B_\eta(u_1^+ \!\!\to \! U^{\rm as}) \, s_1^+ \, ; \, S_1^- = B_\eta(u_1^- \!\!\to \! U^{\rm as}) \, s_1^- \,.
\ee
They involve a flat Poincar\'e-Minkowski metric $\eta_{\mu \nu}$, and the common asymptotic value of $U$ at $\pm \infty$ that we
denoted as $U^{\rm as}$.

To end this Section, let us display the (easily computed) explicit value of the general boost linear operator
 $B_\eta(u \!\!\to \! v)$ in Minkowski space\footnote{Actually, if we interpret $u_\mu$ as $g_{\mu \nu} u^\nu$, 
 $u \cdot v= g_{\mu \nu} u^\mu v^\nu$, etc., the
 formulas below hold in a curved spacetime.}
(for two future-directed timelike unit 4-vectors $u$ and $v$). It reads (suppressing, for brevity, the subscript $\eta$)
\bea \label{boost1}
 && \! \! \!\left[B(u \to v)\right]^\mu{}_\nu = 
\delta^\mu{}_\nu + (u^\mu-v^\mu)u_\nu  \nonumber \\ 
&+&\frac{1}{1-u\cdot v}(u^\mu+v^\mu)[v_\nu +(u\cdot v)u_\nu] \,,
\eea
or, equivalently,
\bea
 &&\left[B(u \to v)\right]^\mu{}_\nu =
\delta^\mu{}_\nu \nonumber \\  \! \! \!  &+& \frac{1}{1-u\cdot v}\left[u^\mu u_\nu +v^\mu v_\nu +  u^\mu v_\nu -  (1-2 u\cdot v) v^\mu u_\nu  \right] . \nonumber \\
\eea
The Lorentz-boost matrix $ B^\mu{}_\nu \equiv \left[B(u \to v)\right]^\mu{}_\nu$ satisfies 
$\eta_{\mu \mu'} B^\mu{}_\nu  B^{\mu'}{}_{\nu'}= \eta_{\nu \nu'}$ and $B^\mu{}_\nu \, u^\nu=v^\mu$.
In addition, when acting on a vector $X^\mu$ orthogonal to $u^\mu$  (i.e. $u \cdot X=0$), it transforms it into the vector
\be
 B^\mu{}_\nu \, X^\nu= X^\mu + \frac{(v \cdot X)}{1- (u \cdot v)} \left(  u^\mu + v^\mu \right) \,,
\ee
which is orthogonal to $v^\mu$. One also easily checks that the inverse matrix of $\left[B(u \to v)\right]^\mu{}_\nu$
is $\left[B(v \to u)\right]^\mu{}_\nu$. Beware, however, that the naively expected composition rule 
$ B(v \to w) \circ B(u \to v)= B(u \to w)$ is {\it only valid} when the third vector $w$ belongs to the 2-plane spanned by $u$ and $v$.
[This is linked to the well-known non-commutativity of non-parallel boosts.]

In the following, we shall work, as is standard in EOB theory, in the center-of-mass (c.m.) frame of the binary system. This corresponds to
choosing
\be
U^{\rm as}= \left[\frac{p_1+p_2}{{\mathcal E}_{\rm real}}\right]^+= \left[\frac{p_1+p_2}{{\mathcal E}_{\rm real}}\right]^- \,,
\ee
where we used the fact that we are considering a conservative dynamics. Here, $ {\mathcal E}_{\rm real}$ denotes the total energy
of the binary system (including the rest-mass energy) in the c.m. frame, which is precisely defined as being the (Minkowski) norm of the
asymptotic value of $p_1+p_2$:
\be
{\mathcal E}_{\rm real}^2 = - (p_1^++ p_2^+)^2 =  - (p_1^-+ p_2^-)^2  \,.
\ee

\section{Spin holonomy}

By combining Eqs. \eqref{scatteringoperator}, \eqref{S1as} above, we obtain the linear map between the two asymptotic values (at $\pm \infty$)
of the spatial spin vector of the first particle, namely
\be
S_1^+= R_1 \, S_1^-  \,,
\ee
where
\be \label{R1}
R_1= B_\eta(u_1^+ \!\!\to \! U^{\rm as}) \, \La_1 \, \left[ B_\eta(u_1^- \!\!\to \! U^{\rm as}) \right]^{-1} \,,
\ee
or, equivalently,
\be \label{R1bis}
R_1= B_\eta(u_1^+ \!\!\to \! U^{\rm as}) \, \La_1 \,  B_\eta(U^{\rm as} \!\!\to \! u_1^-) \,,
\ee
with a similar result for the second particle. Note that $R_1$ is thereby given as the (matrix) product of three
$ 4\times4$ matrices.

The linear operator $R_1$ is easily seen to leave $U^{\rm as}$ invariant: $ R_1 U^{\rm as}=U^{\rm as}$. In addition,
as all the linear maps involved in Eq. \eqref{R1} preserve the (Minkowski) length, and as $R_1$ transforms $ S_1^-$
into $S_1^+$ (both spin vectors living in the three-space orthogonal to $U^{\rm as}$),
we conclude that the linear map $R_1$,
that we shall call the {\it spin holonomy} of $\cL_1$, is an  $SO(3)$ rotation acting within the three-space orthogonal to $U^{\rm as}$
(in the asymptotic Minkowski space).

We can express the spin holonomy, Eqs. \eqref{R1},  \eqref{R1bis}, entirely in terms of the incoming asymptotic values by replacing $u_1^+$ by $ \La_1  u_1^-$,
so that
\be \label{R1ter}
R_1= B_\eta(\La_1 u_1^- \!\!\to \! U^{\rm as}) \, \La_1 \, \left[ B_\eta(u_1^- \!\!\to \! U^{\rm as}) \right]^{-1}\,.
\ee
This is the core theoretical result of our new approach.

\section{Post-Minkowskian computation of the scattering and spin holonomies}

In order to explicitly evaluate the scattering holonomy \eqref{La11}, and the corresponding spin holonomy \eqref{R1ter}, we need 
to use a perturbative approach to their computation.
We could use PN theory, but as PN theory has already been used to derive in a different way the spin-orbit
couplings we are interested in \cite{Damour:2008qf,Nagar:2011fx,Barausse:2011ys}, we shall instead use PM theory 
\cite{Westpfahl:1979gu,Portilla:1979xx,Portilla:1980uz,Bel:1981be,Damour:1981bh,Westpfahl:1985,Westpfahl:1987,Ledvinka:2008tk} 
to show how it allows one to derive new results, valid to all orders in $v/c$.

At the first PM (1PM) order, i.e. when solving the linearized Einstein equations in harmonic coordinates, 
the metric generated by our binary system is of the form $g_{\mu \nu} = \eta_{\mu \nu} + h_{\mu \nu} +O(G^2)$,
with 
\be
h_{\mu \nu}= h_{1 \, \mu \nu} + h_{2 \,\mu \nu}\,,
\ee
where $ h_{1 \, \mu \nu}$ is generated by $\cL_1$ and $ h_{2 \, \mu \nu}$  by $\cL_2$. 
It is well-known that, a linear order in $G$, one can neglect self-force effects (see, e.g., \cite{Bel:1981be}). 
Therefore, when computing the scattering holonomy
along $\cL_1$, the (regularized) metric to be used for computing the parallel transport is simply the contribution
$ h_{2 \, \mu \nu}$  from $\cL_2$.  In principle, $ h_{2 \, \mu \nu}$
contains both a contribution proportional to $m_2$ and one proportional to $s_2$. But, as we are interested in linear-in-spin effects,
it is enough to include in $ h_{2 \, \mu \nu}$ only the contribution generated by $m_2$. Finally, we deal (along $\cL_1$)
with the metric
\be
g_{1 \,\mu \nu} = \eta_{\mu \nu} + h_{2 \, \mu \nu} \, ; \, {\rm with} \,  h_{2 \, \mu \nu} \propto G m_2 \,.
\ee  
The corresponding value, $\omega_1$, of the connection is of order $O(G)$. Working to first order in $G$, we then easily get
the value of the scattering holonomy as
\be
\La_1= 1 -\int_{-\infty}^{+\infty} \omega_1 + O(G^2) \,,
\ee
or, more explicitly,
\be
\La^\mu_{1 \, \nu}= \delta^\mu_{\; \nu}  -\int_{-\infty}^{+\infty} \omega^\mu_{1 \, \nu} + O(G^2)\,.
\ee
Here, 
\be
\omega^\mu_{1 \, \nu} = \eta^{\mu \sigma} \omega_{1 \, \mu \nu} +O(G^2)\,,
\ee
with (going back, for a moment, to an exact expression)
\be
\omega_{1 \, \mu \nu} \equiv g_{\mu \sigma} \G^\sigma_{\; \; \nu \lambda} dx^{\lambda}= \frac12 \left( \D_\nu h_{\mu \lambda } - \D_\mu h_{\nu \lambda} + \D_\lambda h_{\mu \nu}  \right) dx^{\lambda}.
\ee
Note that the last term in this expression for $\omega_{1 \, \mu \nu} $ is the total derivative $\frac12 dh_{\mu \nu}(x)$
(evaluated along $\cL_1$),
 which integrates to zero because of the
asymptotic flatness. This leaves an expression for $\omega_{1 \, \mu \nu}$ which is antisymmetric in $\mu \nu$, and is a curl.
We then get
\be \label{La1PM}
\La^\mu_{1 \, \nu}= \delta^\mu_{\; \nu}  + \eta^{\mu \sigma} \theta_{1 \, \sigma \nu} + O(G^2)\,,
\ee
with
\be \label{theta1h}
\theta_{1 \, \mu\nu}=  + \frac12  \int_{-\infty}^{+\infty} \left( \D_\mu h_{2 \,\nu \lambda } - \D_\nu h_{2 \,\mu \lambda} \right) dx^{\lambda}  + O(G^2)\,.
\ee
Note the antisymmetry of $\theta_{1 \, \mu\nu}= - \theta_{1 \, \nu\mu}$, as expected for an infinitesimal ($O(G)$)  Lorentz transformation.

Under an $O(G)$ coordinate transformation $x'^\mu=x^\mu- \xi^\mu$ (which changes $h_{2 \,\mu \nu}$ into
$h_{2 \,\mu \nu} + \D_\mu \xi_\nu + \D_\nu \xi_\mu + O(G^2)$), it is easily seen that $\theta_{1 \, \mu\nu}$ gets the additional term
\bea
\Delta \theta_{1 \, \mu\nu}&=& \frac12  \int_{-\infty}^{+\infty} \D_\lambda \left( \D_\mu \xi_{\nu} - \D_\nu \xi_{\mu} \right) dx^{\lambda}  + O(G^2) \nonumber \\
&=& \frac12 \left[ \D_\mu \xi_{\nu} - \D_\nu \xi_{\mu} \right]_{-\infty}^{+\infty} \,.
\eea
This vanishes when considering a coordinate transformation $\xi^\mu$ which decays at infinity. [More precisely, it suffices that
$\D_\mu \xi_{\nu} - \D_\nu \xi_{\mu}$ decays at infinity.] This directly confirms the (geometrically evident)
gauge invariance of $\theta_{1 \, \mu\nu}$.

It is straightforward to compute the 1PM-accurate scattering holonomy $\theta_{1 \, \mu\nu}$ from the well-known
1PM metric generated by $\cL_2$. Using, for instance, the results given in Appendices A and B of Ref. \cite{Bel:1981be},
and using the simplifying fact that, at this order, we can consider that $\cL_2$ is a straight worldline (with tangent $u_2=u_2^-$),
we have (at an arbitrary field point $x^\mu$)
\be \label{h2}
h_{2 \,\mu \nu}(x) = 2 \frac{G m_2}{R_2} \left( 2 \, u_{2 \, \mu} u_{2 \, \mu} + \eta_{\mu\nu} \right)\,,
\ee
where $R_2=R_2(x)$ denotes the Poincar\'e-invariant orthogonal distance between the field point $x$ and the straight worldline 
$\cL_2$ (= $\cL_2^-$, at this order). Explicitly, $R_2(x) = | x - x_2^\perp(x)|$, where $x_2^\perp(x)$ denotes the foot of the perpendicular\footnote{One uses 
here the Poincar\'e-Minkowski geometry.} of the field point $x$ on the line $\cL_2$. 
 [In the case where one must take into account the $O(G)$ curvature of $\cL_2$ the expression of 
$h_{2 \,\mu \nu}(x)$ should involve the half-sum of  retarded and advanced tensor potentials generated by $\cL_2$.]
The Poincar\'e-Minkowski geometrical situation underlying the definition of $x_2^\perp(x)$ is illustrated in Fig. 2.

\begin{figure}
\includegraphics[scale=0.99]{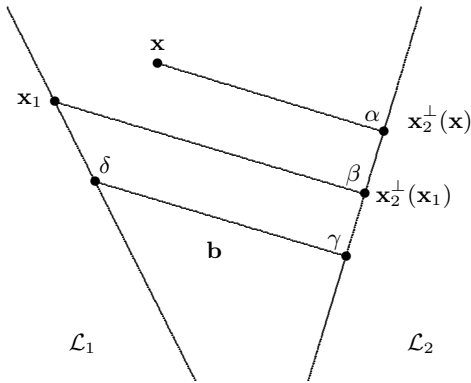}
\caption{\label{fig:2}  Geometrical configuration (in Minkowski spacetime) underlying the computation of the scattering holonomy at
the 1PM order. The 1PM metric generated by $\cL_2$ is computed at the field point $x$ (which is then made to move all
along $\cL_1$). The point $x_2^\perp(x) \in \cL_2$ is the foot of the perpendicular drawn from the field point $x$ to $\cL_2$. The segment
$b$ (oriented from $\cL_2$ towards $\cL_1$) denotes the
four-vectorial impact parameter of $\cL_1$ wrt $\cL_2$. All the labelled angles ($\alpha, \beta,\gamma,\delta$) indicate orthogonal
incidences (in the Minkowski sense).}
\end{figure}

The partial derivative wrt $x$ of $h_{2 \,\mu \nu}(x)$ is then (using again $\dot u_2=O(G)$)
\be \label{Dh2}
\D_\lambda h_{2 \,\mu \nu}(x)= - 2 \frac{G m_2 R_{2 \, \lambda}}{R_2^3} \left( 2 \, u_{2 \, \mu} u_{2 \, \nu} + \eta_{\mu\nu} \right)+O(G^2)\,,
\ee
where $R_2^{\lambda} = x^{\lambda} - x_{2 \,\perp}^{\lambda}(x)$ is the 4-vector connecting the perpendicular foot $x_2^\perp(x)$ to $x$.

We must then evaluate along $\cL_1$ (i.e. replace $x\to x_1(\tau_1)$) the combination of partial derivatives of  $h_{2 \,\mu \nu}(x)$ 
entering \eqref{theta1h}, and integrate over $\cL_1$ using $dx^{\lambda} = u_1^{\lambda} d \tau_1$. This computation involves
the easily evaluated integral (where  $x_1 \equiv x_1(\tau_1)$)
\be \label{int1}
\int_{\cL_1} d \tau_1 \frac{x_1^{\mu} - x_{2 \,\perp}^{\mu}(x_1)}{|x_1 - x_{2 \,\perp}(x_1)|^3} = \frac{2}{\sqrt{(u_1\cdot u_2)^2-1}}\,  \frac{b^\mu}{b^2}\,.
\ee
Here, $b^\mu$ denotes the Poincar\'e-invariant four-vectorial impact parameter of $\cL_1$ wrt $\cL_2$. We mean by this the value of the
four-vector $x_1^{\mu}(\tau_1) - x_{2 \,\perp}^{\mu}\left(x_1(\tau_1)\right)$ at the moment of closest approach, i.e. for the
value of $\tau_1$ minimizing $|x_1 - x_{2 \,\perp}(x_1)|$. This vectorial impact parameter is simply characterized as being
the connecting four-vector $x_1^\mu - x_2^\mu$ that is perpendicular  {\it both} to $u_1$ and to $u_2$ (see Fig. 2). 
It is convenient to choose the origins of the proper-time parameters along $\cL_1$ and $\cL_2$ as the end points of $b^\mu$ 
(labelled by $\delta$ and $\gamma$ in Fig. 2), so that we can write the equations of $\cL_1$ and $\cL_2$ as: 
$x_1^\mu(\tau_1) = x_1^\mu(0) + u_1^\mu \tau_1$, and $x_2^\mu(\tau_2) = x_2^\mu(0) + u_2^\mu \tau_2$, 
with $ x_1^\mu(0) -  x_2^\mu(0)= b^\mu$.
[The vectorial impact parameter of $\cL_2$ wrt $\cL_1$ is simply $ x_2^\mu(0) -  x_1^\mu(0) = - b^\mu$.]

Inserting Eqs. \eqref{Dh2} and \eqref{int1} into Eq. \eqref{theta1h} yields the explicit result
\be \label{theta1PM1}
(\theta_1)_{\mu\nu} = -\frac{4 \,Gm_2 \, b_{[\mu }}{ b^2 \sqrt{(u_1 \cdot u_2)^2-1}}(2 (u_1\cdot u_2)\,   u_{2 \,\nu]} + u_{1 \,\nu]})\,,
\ee
where $a_{[\mu} b_{\nu]} \equiv \frac12( a_\mu b_\nu - a_\nu b_\mu)$.

Note that this result features the {\it past}-directed timelike vector
\be
V_2^\mu \equiv 2 (u_1 \cdot u_2) \, u_2^\mu + u_1^\mu ; ({\rm satisfying} \,  V_2 \cdot V_2 = -1) \,;
\ee
namely
\be\label{theta1PM2}
(\theta_1)_{\mu\nu} = -\frac{2 \,Gm_2 }{ b^2 \sqrt{(u_1 \cdot u_2)^2-1}} \left( b_\mu  V_{2 \, \nu} -  b_\nu  V_{2 \, \mu}\right)\,.
\ee
Note that while $b^\mu$ is {\it antisymmetric} under the $1 \leftrightarrow 2$ exchange,  $V_2$ is {\it asymmetric}. 

An alternative way of deriving  the latter result  is to first compute the specialized value of $(\theta_1)_{\mu\nu}$ in the
rest-frame of $\cL_2$ (which means computing the scattering holonomy of a test particle of mass $m_1$ moving in the background of
a linearized Schwarzschild metric of mass $m_2$), and to then re-express the result in a manifestly Poincar\'e-invariant way.

As a first check on the 1PM result, Eqs. \eqref{La1PM}, \eqref{theta1PM1}, \eqref{theta1PM2}, let us compute the orbital part of
the scattering, i.e. the change in the direction of $\cL_1$:
\be
\Delta u_1 \equiv u_1^+   -  u_1^- = \La_1  u_1^- -  u_1^- \,.
\ee
In the following, we will always think of $\La_1 = \La^\mu_{1 \, \nu}$ as a matrix (an endomorphism of Minkowski spacetime),
and use the simplified notation
\be
\La_{1 }= 1  + \theta_1+ O(G^2) \,,
\ee
where $1$ denotes the unit matrix (i.e. $\delta^\mu_{\,\, \nu}$),
while $\theta_1$ will denote the matrix (endomorphism)  $ \theta^\mu_{1 \,  \nu} = \eta^{\mu \sigma} \theta_{1 \, \sigma \nu}$.
With this notation the change
\be
\Delta u_1 = \theta_1 \, u_1^- ; \, {\rm i.e.} \, \,\Delta u_1^\mu  = \theta^\mu_{1 \,  \nu} \, u_1^{- \, \nu}\,,
\ee
is easily computed from the explicit result \eqref{theta1PM1}, \eqref{theta1PM2} to be
\be
\Delta u_1^\mu =  -2G\,m_2 \frac{  2 (u_1 \cdot u_2)^2 -1 }{  \sqrt{(u_1 \cdot u_2)^2-1}} \frac{ b^\mu}{b^2}\,,
\ee
i.e., with $p_1^\pm= m_1 u_1^\pm$
\be
\Delta p_1^\mu =  -2G\,m_1m_2 \frac{  2 (u_1 \cdot u_2)^2 -1 }{  \sqrt{(u_1 \cdot u_2)^2-1}} \frac{ b^\mu}{b^2}\,.
\ee
Note that the right-hand side (rhs) of the latter result is antisymmetric under the $1 \leftrightarrow 2$ exchange:
$\Delta p_1= - \Delta p_2$, i.e.  conservation of the total four-momentum:
$p_1^- + p_2^- = p_1^+ + p_2^+$.
The result for $\Delta p_1$ is easily checked to be in agreement with the known 1PM scattering angle results; 
see, e.g., Eqs. (55)-(58) in Ref. \cite{Damour:2016gwp}.

With this definition, and using matrix notation (for endomorphisms) we find that the 1PM estimate of the spin holonomy, Eq. \eqref{R1ter},
reads
\begin{eqnarray}
R_1 &=& B_\eta((1+\theta_1) u_1^- \!\!\to \! U^{\rm as}) \, (1+\theta_1) \, \left[ B_\eta(u_1^- \!\!\to \! U^{\rm as}) \right]^{-1}\nonumber\\
&+& O(G^2).
\end{eqnarray}
Expanding it to first order in $G$ yields an infinitesimal $SO(3)$ rotation of the form
\be
R_1=1+ \rho_1 + O(G^2)\,,
\ee
where
\bea \label{R1PM}
 \rho_1&=&  B_\eta( u_1^- \!\!\to \! U^{\rm as}) \theta_1 \left[ B_\eta(u_1^- \!\!\to \! U^{\rm as}) \right]^{-1} \nonumber \\
 &+& \left( (\theta_1 u_1^-) \frac{\D}{\D u_1^-} B_\eta( u_1^- \!\!\to \! U^{\rm as}) \right) \left[ B_\eta(u_1^- \!\!\to \! U^{\rm as}) \right]^{-1} \nonumber \\
 &+& O(G^2) \,,
\eea
belongs to the Lie algebra of spatial rotations in the 3-plane orthogonal to $U^{\rm as}$.
Note the second contribution coming from expanding $u_1^+= (1+\theta_1) u_1^-$ in the argument of the first boost transformation.
[In this contribution $ (\theta_1 u_1^-) \frac{\D}{\D u_1^-}=  {\theta_1}^\mu_{\, \,\nu} u_1^{- \nu}\frac{\D}{\D u_1^{-  \mu}}$ denotes a directional derivative.]
The result \eqref{R1PM} is entirely expressed in terms of $\theta_1$ and of the ingoing four-velocity. Moreover, $\theta_1$ can also
be expressed (modulo $O(G^2)$) entirely in terms of incoming data. [To ease the notation we henceforth suppress the superscripts $-$
on the incoming data.]

As the antisymmetric tensor $\rho_{1 \, \mu \nu} = \eta_{\mu \sigma} \rho_{1 \, \, \nu}^{\, \sigma}$ is orthogonal to $U^\mu_{\rm as}$,
and is easily seen (from the explicit expression of the boost operator \eqref{boost1} and of $\theta_1$) to be algebraically
constructed from the 4-vectors $b$, $u_1=p_1/m_1$ and $u_2=p_2/m_2$, it must involve (besides $b$ which, being orthogonal
to $u_1$ and $u_2$, is already orthogonal to $U_{\rm as}$) the projections of $p_1$ and $p_2$ orthogonally to $U_{\rm as}$, i.e. the following
Minkowski-covariant version of the c.m. momentum:
\be
\pcm^\mu\equiv\left( \delta^\mu_{\, \,\nu} + U^\mu_{\rm as} U^{\rm as}_{\, \nu}  \right) p_1^\nu = - \left( \delta^\mu_{\, \,\nu} + U^\mu_{\rm as} U^{\rm as}_\nu  \right) p_2^\nu \,. 
\ee
Therefore, $\rho_{1 \, \mu \nu}$ must be of the form
\be \label{rho=CL}
\rho_{1 \, \mu \nu} = C_1 \, L_{\mu \nu} \,,
\ee
where we defined the following covariant version of the orbital angular momentum
\be
L_{\mu \nu} = b_\mu \, p_{{\rm c.m.} \, \nu} -  b_\nu \,p_{{\rm c.m.} \, \mu} \,.
\ee

A straightforward calculation of $\rho_{1 \, \mu \nu}$ obtained by inserting Eq. \eqref{theta1PM2} into Eq. \eqref{R1PM}, 
indeed leads to the form \eqref{rho=CL} with
\be
C_1=-\frac{2 G h}{b^2  (w-1)^{1/2} (w+1)^{3/2}} \left({\sf c} + \frac{m_2}{m_1} {\sf c}_* \right) \,.
\ee
Here, we used the notations,
\be \label{w}
w \equiv - (u_1 \cdot u_2) \, ; \, ({\rm so \,\, that} \, w >0)
\ee
for the relative Lorentz $\gamma$ factor between $\cL_1$ and $\cL_2$, 
\be
h= \frac{{\mathcal E}_{\rm real}}{M} = \sqrt{1+2\nu (w-1)},
\ee
for a dimensionless measure of the total energy of the system, and
\be
{\sf c}= \frac{(1+2w)(h+2w)-1}{1+h} \,,
\ee
\be
{\sf c}_*=1+2w \,.
\ee

\section{EOB computation of the spin holonomies} \label{sec3}

Let us now see how one can relate the gauge-invariant (three-dimensional) spin holonomies $R_1$ and $R_2$ to the
spin-orbit couplings entering the EOB Hamiltonian.  We recall that the (real) EOB Hamiltonian has the form
\be \label{Heob}
H({\mathbf R}, {\mathbf P}, \vS_1, \vS_2)=M \sqrt{1+2\nu \left(\frac{H_{\rm eff}}{\mu }-1 \right)}\,,
\ee
where, at linear order in the spins, the effective EOB Hamiltonian $H_{\rm eff}$ reads
\begin{eqnarray}
\label{Heff}
H_{\rm eff}&=&  \sqrt{A\left( \mu^2  +{\mathbf P}^2 +\left(\frac{1}{B}-1\right)P_R^2+Q \right)}\nonumber\\
&+& \frac{G}{ R^3} \left(g_S \mathbf{L} \cdot \mathbf{S} + g_{S_*} \mathbf{L} \cdot \mathbf{S_*} \right) \,.
\end{eqnarray}
Here the functions $A(R)$ and $B(R)$ parametrize the effective metric
\be\label{geff}
ds_{\rm eff}^2=-A(R)c^2 dT_{\rm eff}^2+B(R) dR^2 +R^2 (d \theta^2 + \sin^2 \theta \, d\phi^2)\,,
\ee
$P_R$ denotes the EOB radial momentum, while
\be
{\mathbf P}^2\equiv P_R^2+\frac{{\mathbf L}^2}{R^2}\,, 
\ee
where ${\mathbf L}$ denotes the EOB orbital angular momentum 
\be
{\mathbf L}={\mathbf R}\times {\mathbf P}\,.
\ee
[Here, and below, we use standard vectorial notation for various EOB vectorlike objects.]
In addition, $Q$ in Eq. \eqref{Heff} represents a post-geodesic (Finsler-type) contribution which is at least quartic in momenta. Finally, $\mathbf{S}$ and $\mathbf{S_*}$ denote the following symmetric combinations of the two spin vectors
\be
\mathbf{S}= \mathbf{S_1} + \mathbf{S_2} \, ; \, \mathbf{S_*}= \frac{m_2}{m_1} \mathbf{S_1} + \frac{m_1}{m_2} \mathbf{S_2} \,,
\ee
while $g_S$ and  $g_{S_*} $ are some corresponding  gyro-gravitomagnetic ratios (introduced in \cite{Damour:2008qf}) which are the focus of the present work. 

 As shown in Ref.\cite{Damour:2016gwp}, at 1PM order, the post-geodesic term $Q$ is zero, while the effective metric \eqref{geff} is simply
equal to a linearized Schwarzschild metric of mass $M=m_1+m_2$, i.e.
\bea
A(R)&=& 1 - 2 \frac{GM}{R} +O(G^2)\; ; \nonumber\\ 
B(R)&=& 1 + 2 \frac{GM}{R} +O(G^2) \; ; \nonumber\\
 Q&=&O(G^2).
\eea
The two (dimensionless) gyro-gravitomagnetic ratios $g_S$ and  $g_{S_*} $ are functions of ${\mathbf P}^2/\mu^2$, $ P_R^2/\mu^2$, 
$\nu$ and $u\equiv GM/R$.
Their current knowledge is the following: (i) their PN expansion is known
to the next-to-next-to-leading-order level \cite{Damour:2008qf,Nagar:2011fx,Barausse:2011ys}; (ii) the test-mass limit ($\nu \to 0$) of the
second  gyro-gravitomagnetic ratio $g_{S_*} $ is known exactly \cite{Barausse:2009aa,Barausse:2009xi}; and (iii) recent self-force computations have allowed
one to compute, to high PN order, the circular, and next-to-circular, contributions to $g_S$ and  $g_{S_*} $, 
see \cite{Bini:2014ica,Bini:2015xua,Kavanagh:2017wot}, and references therein.

Here, we consider the PM expansions of  $g_S$ and  $g_{S_*} $, i.e., keeping in mind that $u\equiv GM/R = O(G)$, their expansions in
powers of $u$:
\bea
g_S({\mathbf P}^2, P_R^2, u) &=& g^{1\rm{PM}}_S({\mathbf P}^2, P_R^2) + u\,  g^{2\rm{PM}}_S({\mathbf P}^2, P_R^2)\nonumber\\
& +& O(u^2) \nonumber \\
g_{S_*}({\mathbf P}^2, P_R^2, u) &=& g^{1\rm{PM}}_{S_*}({\mathbf P}^2, P_R^2) + u\,  g^{2\rm{PM}}_{S_*}({\mathbf P}^2, P_R^2)\nonumber\\
& +& O(u^2)\,.
\eea
We have labelled the momentum-dependent, but $u$-independent, leading-order contributions  as being of the first PM order because
they enter the effective Hamiltonian \eqref{Heff} multiplied by the prefactor $G/R^3$.

The equation of motion of the spin vector $\vS_1$ deduced from the effective Hamiltonian \eqref{Heff} reads
\be
\frac{d \vS_1}{d \Tf} = {\mathbf \Omega}_1^{\rm eff} \times \vS_1 \,,
\ee
where
\be
{\mathbf \Omega}_1^{\rm eff} = \frac{G}{R^3} {\mathbf L} \left( g_S + \frac{m_2}{m_1}  g_{S_*} \right) \,,
\ee
and where $\Tf$ is the ``effective time" entering \eqref{geff}, i.e. the evolution parameter associated with the 
dynamics defined by $H_{\rm eff}$. It differs from the real (coordinate) time $T$ by a factor $dH/dH_{\rm eff}$ \cite{Buonanno:1998gg}.
For our present purpose, we do not need to consider the evolution wrt the real time $T$.  Actually, it is convenient to rewrite the
evolution equation for $\vS_1$ in the differential form 
\be \label{eomS1}
d \vS_1 = {\boldsymbol \omega}_1^{\rm eff} \times \vS_1 \,,
\ee
where
\bea
 && {\boldsymbol \omega}_1^{\rm eff} = {\mathbf \Omega}_1^{\rm eff} d\Tf \nonumber \\
 &=&  \frac{G}{R^3} {\mathbf L} \left( g_S + \frac{m_2}{m_1}  g_{S_*} \right) \, d\Tf \,.
\eea
The notation ${\boldsymbol \omega}_1^{\rm eff}$ should not be confused with the notation $\omega_1$ used above.
Both quantities describe the same physics (the infinitesimal rotation of the spin), but they live in different frameworks.

The evolution equation for $\vS_2$ is obtained by exchanging everywhere the labels $1 \leftrightarrow 2$.

The spin holonomy of $\vS_1$ computed in EOB theory is simply obtained by integrating the linear evolution equation \eqref{eomS1}, i.e.
\be
R_1^{\rm EOB} = T e^{\int_{-\infty}^{+\infty}  {\boldsymbol \omega}_1^{\rm eff} \times}
\ee
where  $ {\boldsymbol \omega}_1^{\rm eff} \times $ is viewed as a linear operator (actually an infinitesimal rotation) acting 
(as ${\mathbf v} \to {\boldsymbol \omega}_1^{\rm eff} \times {\mathbf v}$) in a three-dimensional Euclidean vector space .

As $ {\boldsymbol \omega}_1^{\rm eff}$ is of order $G$, the 1PM approximation to the EOB spin holonomy $R_1^{EOB} $ is simply
\be
R_1^{\rm EOB} = 1 + {\boldsymbol \theta}_1 \times + O(G^2)\,,
\ee
where the infinitesimal (i.e. $O(G)$) vectorial rotation angle experienced by $\vS_1$ during the entire scattering is
\bea \label{theta1}
&& {\boldsymbol \theta}_1 = \int_{-\infty}^{+\infty} {\boldsymbol \omega}_1^{\rm eff}  \nonumber \\
 &&= \int_{-\infty}^{+\infty}   \frac{G}{R^3} {\mathbf L} \left( g^{1\rm{PM}}_S \!({\mathbf P}^2, P_R^2) + \frac{m_2}{m_1}  g^{1\rm{PM}}_{S_*} \!({\mathbf P}^2, P_R^2) \right) d\Tf \nonumber
\eea
To turn this result into a fully explicit integral we just need to express $d\Tf $ in terms of dynamical variables. This is obtained by using
the equation of motion of $R$ deduced from $H_{\rm eff}$, namely
\be \label{eomR}
\frac{d R}{d \Tf} = \frac{\D H_{\rm eff}}{\D P_R} = \frac{A}{B} \frac{P_R}{H_{\rm eff}} \,,
\ee
i.e.
\be
d \Tf =  \frac{B}{A} H_{\rm eff} \frac{dR}{P_R}\,.
\ee
Actually, as $ {\boldsymbol \theta}_1$ is of order $G$, we can replace $\Tf$ by its 0PM ($O(G^0)$) approximation, i.e. use 
$A \approx 1$, $B\approx 1$, and
\be
H_{\rm eff}=  \sqrt{ \mu^2  +{\mathbf P}^2 } + O(G) = \sqrt{ \mu^2 + P_R^2+\frac{{\mathbf L}^2}{R^2} }+O(G)\,.
\ee
Moreover, in this approximation the orbital angular momentum ${\mathbf L}$ entering the integral $ {\boldsymbol \theta}_1$
is constant. We can then use $R$ as integration variable, and compute $P_R$ as a function of $R$ and of the 0PM constants of
the motion $E_{\rm eff} = H_{\rm eff}$ and ${\mathbf L}$ via
\be \label{PR}
P_R = \pm \sqrt{E_{\rm eff}^2 - \mu^2 - \frac{{\mathbf L}^2}{R^2} }\,.
\ee
Finally, we have
\bea
{\boldsymbol \theta}_1 &=& G \int_{-\infty}^{+\infty}    \frac{{\mathbf L}}{R^3}  \left( g^{1\rm{PM}}_S \!({\mathbf P}^2, P_R^2) + \frac{m_2}{m_1}  g^{1\rm{PM}}_{S_*} \!({\mathbf P}^2, P_R^2) \right) \nonumber\\
&\times& \frac{E_{\rm eff} dR}{ P_R} \,.
\eea
Here, we kept the time-integration limits ${-\infty}, {+\infty}$ on the integral sign to indicate the physical integration
along the entire motion. The corresponding variation of $R$ goes from ${+\infty}$ to $R_{\rm min}$ (with a negative
choice for the root defining $P_R$ in Eq. \eqref{PR}), and then from
$R_{\rm min}$ back to ${+\infty}$ (with the positive root for $P_R$ in Eq. \eqref{PR}). The net effect is that 
${\boldsymbol \theta}_1$ can be written as being {\it twice} the integral from $R_{\rm min}$ back to ${+\infty}$,
with a positive value for $P_R$.

Let us now remark that the final result for the integrated $O(G)$ spin rotation ${\boldsymbol \theta}_1$ involves the
two gyro-gravitomagnetic ratios  $g^{1\rm{PM}}_S \!({\mathbf P}^2, P_R^2)$ and $g^{1\rm{PM}}_{S_*} \!({\mathbf P}^2, P_R^2)$
only through the integrals $\int R^{-3} g_S d\Tf$ and $\int R^{-3} g_{S_*} d\Tf$. Therefore, the physically relevant integrated
spin rotation is left invariant under independent $O(G)$ ``gauge transformations" of $g^{1\rm{PM}}_S \!({\mathbf P}^2, P_R^2)$ and $g^{1\rm{PM}}_{S_*} \!({\mathbf P}^2, P_R^2)$ of the type
$g'_S=g_S/R^3 + df_S(P_R,R,L)/d\Tf$, and $g'_{S_*}=g_{S_*}/R^3 + df_{S_*}(P_R,R,L)/d\Tf$,
involving two arbitrary phase-space functions $f_S(P_R,R,L), f_{S_*}(P_R,R,L)$. [The effective-time derivatives have to be
interpreted as  Poisson bracket $ \{ f_S, H_{\rm eff} \}$, $ \{ f_{S_*}, H_{\rm eff} \}$, evaluated with the
effective Hamiltonian.] It is easily checked that this gauge-freedom in the definitions of $g_S$ and $g_{S_*}$
is the 1PM analog of the $O(1/c^4)$ PN gauge-freedom associated with the canonical transformation Eq. (3.4) 
in Ref. \cite{Damour:2008qf} (and extended to the next PN order in \cite{Nagar:2011fx,Barausse:2011ys}). As
remarked in footnote 9 of \cite{Damour:2008qf}, this gauge-freedom comes from the arbitrariness in the choice of
a local frame to measure the spatial spin vectors. In the PN framework, this arbitrariness is parametrized, at each
PN order by a finite numer of parameters (see below). In our present PM framework, the arbitrariness is larger because
it involves, at each PM order (say the $n$PM level), two functions,  which, as is easily seen, can be taken of the form
$f_S= P_R R^{-2}{\hat f}_S(P_R^2, L^2/R^2) u^{n-1}$, and $f_{S_*}= P_R R^{-2}{\hat f}_{S_*}(P_R^2, L^2/R^2) u^{n-1}$.
In particular, at the 1PM order ($n=1$), it is easy to check that this large functional freedom can arbitrarily change the
$P_R$ dependence of $g^{1\rm{PM}}_S \!({\mathbf P}^2, P_R^2)$ and $g^{1\rm{PM}}_{S_*} \!({\mathbf P}^2, P_R^2)$.

Ref. \cite{Damour:2008qf}, having in view the application of EOB theory to quasi-circular inspiralling and coalescing 
binary black holes, had suggested to simplify the momentum dependence of $g_S$ and  $g_{S_*} $ by making
them depend (at each PN order) only on $P_R^2$ (and not on ${\mathbf P}^2$). Within our present PM context,
it seems technically more convenient to replace the latter ``DJS gauge", by an ``anti-DJS" gauge where,
at the $n$PM order, $g_S$ and  $g_{S_*} $ only depend (after factoring the PM prefactor $u^{n-1}$) on ${\mathbf P}^2$.
In particular, at the 1PM level, this means using $P_R$-independent gyro-gravitomagnetic ratios of the form
$g^{1\rm{PM}}_S \!({\mathbf P}^2)$ and $g^{1\rm{PM}}_{S_*} \!({\mathbf P}^2)$.

Using such an ``anti-DJS" gauge, it is very simple to compute the integral \eqref{theta1}, because $g_S$ and  $g_{S_*} $
become constant (${\mathbf P}^2$ being constant at the 0PM level) and can be factored out. We have then
\be
{\boldsymbol \theta}_1 =  G \Ef \, {\mathbf L} \, \left(g_S^{\rm 1PM} +\frac{m_2}{m_1} g_{S*}^{\rm 1PM} \right) \, \,{\mathcal I}\,,
\ee
where ${\mathcal I}$ denotes the elementary integral
\begin{eqnarray}
 {\mathcal I}&=& 2 \int_{R_{\rm min}}^\infty \frac{1}{R^3}  \frac{dR}{\sqrt{E_{\rm eff}^2-\mu^2 -\frac{{\mathbf L}^2}{R^2}}}\nonumber\\
&=&\frac{1}{L^2} \int_0^{x_{\rm min}}  \frac{dx}{\sqrt{x_{\rm min} -x}} =\frac{2}{L^2}\sqrt{x_{\rm min}}\,,
\end{eqnarray}
with $x=L^2/R^2$ and $x_{\rm min}= E_{\rm eff}^2-\mu^2$. This yields the final explicit result
\be
{\boldsymbol \theta}_1 = 2 G \Ef \, \sqrt{E_{\rm eff}^2-\mu^2} \frac{\mathbf L}{L^2} \, \left(g_S^{\rm 1PM}\!({\mathbf P}^2) +\frac{m_2}{m_1} g_{S*}^{\rm 1PM}\!({\mathbf P}^2) \right)
\ee

\section{Dictionary between the PM result and the EOB one}

One of the defining features of EOB theory is to be able to identify the spatial spin vectors $\vS_1$, $\vS_2$ entering the
EOB Hamiltonian with the canonical spatial spin vectors entering the real (PN-expanded, or PM-expanded) dynamics,
and also to be able to identify the total c.m. angular momentum of the system $\bf J$ with the EOB total angular
momentum ${\bf L} + \vS_1 +\vS_2$. Because of these identifications, and because the spin vectors become immune
to (asymptotic-flatness respecting) gauge ambiguities when considering incoming and outgoing scattering states, the
dictionary between the real spin holonomy, and its EOB counterpart, is simply the equality
\be \label{dic1}
R_1^{\rm real} = R_1^{\rm EOB} \,.
\ee
When applying this equality in the present case where the lhs is computed within the PM framework
as $R_1^{\rm real}= R_1^{\rm PM}= 1 + \rho_1 +O(G^2)$, 
while the rhs is computed as $R_1^{\rm EOB} = 1 + {\boldsymbol \theta}_1 \times + O(G^2)$,
we simply identify the 3-space orthogonal to $U$ in which $\rho_1$ lives with the 3-space in which ${\boldsymbol \theta}_1$ lives,
so that we get
\be \label{dic2}
\rho_1 =  {\boldsymbol \theta}_1 \times  + O(G^2) \,.
\ee
In terms of tensor components wrt a 3-frame this yields
\be \label{dic3}
\rho_{1 \, ij} = -\theta_{1 \, ij} \,.
\ee
From Eq. \eqref{rho=CL}, the lhs involves 
\be
\rho_{1 \, ij} = C_1 L_{ij} = C_1 \left( b_i \, p_{{\rm c.m.} \, j} -  b_ j\,p_{{\rm c.m.} \, i}  \right) \,.
\ee
But the definition of the 4-vectors $b$ and $\pcm$ have been chosen so that their spatial projections orthogonal to $U$ entering
the above equation are precisely such that $ L_{ij}$ are the components of the c.m. orbital angular momentum. Moreover, the
definition of the EOB formalism is such that the vector ${\mathbf L}$ entering the EOB Hamiltonian can be identified with the
real c.m. angular momentum. [On both sides, PM and EOB, as we talk about linear-in-spin effects, we can treat the orbital dynamics
as if we were treating nonspinning bodies. This allows us to identify\footnote{If we could not neglect spin-orbit corrections
to the orbital dynamics we should examine more carefully the expression of the total angular momentum $J_{\mu \nu}$ in the real dynamics.} $ L_{ij}$ with ${\mathbf L}$.]
We therefore see that the condition  \eqref{dic3} is compatible with the tensor structure of both sides,
and simply leads to an identification of two scalar factors, namely
\be \label{dic4}
- C_1 =\frac{ 2 G \Ef \, \sqrt{E_{\rm eff}^2-\mu^2}} {L^2} \, \left(g_S^{\rm 1PM} +\frac{m_2}{m_1} g_{S*}^{\rm 1PM} \right)\,.
\ee
This gives a condition to determine both  $g^{1\rm{PM}}_S \!({\mathbf P}^2)$ and  $ g^{1\rm{PM}}_{S_*} \!({\mathbf P}^2)$.
Indeed, we have seen above that $-C_1$ had a similar structure $\propto {\sf c} + \frac{m_2}{m_1} {\sf c}_*$,
so that $g^{1\rm{PM}}_S \!({\mathbf P}^2) \propto {\sf c}$ and $ g^{1\rm{PM}}_{S_*} \!({\mathbf P}^2) \propto {\sf c}_*$.

In order to get the explicit values of $g^{1\rm{PM}}_S \!({\mathbf P}^2)$ and  $ g^{1\rm{PM}}_{S_*} \!({\mathbf P}^2)$ we just need
to translate the kinematical PM  quantities $w$, $h$ and $b$ entering $C_1$ into their dynamical EOB counterparts. The EOB dictionary (which has been recently proven
to be valid to all orders in $v/c$ \cite{Damour:2016gwp}) yields the simple links (where it is convenient to define ${\mathbf p} \equiv {\mathbf P}/\mu$):
\be
w= \frac{\Ef}{\mu}= \frac1{\mu} \sqrt{ \mu^2  +{\mathbf P}^2 } + O(G)= \sqrt{ 1  +{\mathbf p}^2 } + O(G)\,,
\ee
and
\be
h = \frac{{\mathcal E}_{\rm real}}{M}=  \sqrt{1+2\nu \left(\frac{\Ef}{\mu}-1\right)} \,.
\ee
In addition, Eqs. (51), (52) and (54) in Ref. \cite{Damour:2016gwp} yield the following links between the impact parameter and PM quantities:
\be
L= b \, \pcm \,,
\ee
and
\be
{\mathcal E}_{\rm real} \, \pcm = {\mathcal D}=\sqrt{(p_1.p_2)^2 - p_1^2 \, p_2^2 } \,,
\ee
i.e.
\be
{\mathcal E}_{\rm real} \, \pcm= m_1 m_2 \sqrt{w^2-1} \,.
\ee
Note that $|{\mathbf P}|$ is not equal to $\pcm$ but rather we have the link
\be
|{\mathbf P}|= h \, \pcm\,.
\ee

Using these links (and remembering the definition $\nu=m_1 m_2/M^2$) we finally derive the 1PM values (
in the anti-DJS gauge) of the
EOB gyrogravitomagnetic factors:
\be
g^{1\rm{PM}}_S =\frac{{\sf c}}{h \, w (1+w)}=\frac{(1+2w)(h+2w)-1}{h\, w(1+w)(1+h)}\,,
\ee
\be
g_{S*}^{\rm 1PM}= \frac{{\sf c}_*}{h \,w (1+w)}=  \frac{1+2w}{h \, w (1+w)}\,,
\ee
i.e., explicitly, introducing the shorthand notation,
\be
w_p \equiv \sqrt{ 1  +{\mathbf p}^2 }\,,
\ee 
\begin{widetext}
\begin{eqnarray}
\label{gsfin}
g^{1\rm{PM}}_S({\mathbf p}^2,\nu) &=& \frac{ \left( 1+2w_p \right)\left(\sqrt{1+ 2 \nu (w_p-1)}+ 2 w_p \right) -1 }{ w_p (1+w_p) \sqrt{1+ 2 \nu (w_p-1)}(1+\sqrt{1+ 2 \nu (w_p-1)}) }\,\\
\label{gssfin}
g^{1\rm{PM}}_{S_*}({\mathbf p}^2,\nu) &=& \frac{ \left( 1+2w_p \right) }{ w_p (1+w_p) \sqrt{1+ 2 \nu (w_p-1)} }\,.
\end{eqnarray}
\end{widetext}
These last two equations are the central new technical results of the present paper.

\section{Comparison with previous results}

Let us now compare our results for   $g^{1\rm{PM}}_S \!({\mathbf p}^2,\nu)$ and  $ g^{1\rm{PM}}_{S_*} \!({\mathbf p}^2,\nu)$ 
to the previously acquired knowledge of these gyrogravitomagnetic ratios. First, let us recall that, at the linear order in spins at which
we are working, and in the anti-DJS gauge we are using,  $g_S $ and $g_{S_*} $ are functions of {\it three} (dimensionless) variables,
namely ${\mathbf p}^2,\nu$ and $u=GM/R$.

In the extreme mass-ratio limit $\nu \to 0$, one knows the exact values of $g_S $ and $g_{S_*} $, namely
\be \label{gsnu0}
\lim_{\nu\to0} g_S\left({\mathbf p}^2,\nu,u \right) =2 \,,
\ee
as deduced in Refs. \cite{Damour:2001tu,Damour:2008qf} from the Kerr metric, and
\be
\label{gssnu0}
\lim_{\nu\to0} g_{S_*} =\frac{1}{1+W}  +\frac{1}{W}\, \frac{2}{1+\frac{1}{\sqrt{1-2u}}}\,,
\ee
as derived in Refs. \cite{Barausse:2009aa,Barausse:2009xi} (and simplified  in Eq. (2.21) of 
\cite{Bini:2015xua}, and Eq. (4.14) of \cite{Kavanagh:2017wot}). Here, $W$ denotes
\be
W=\sqrt{1 + {\mathbf p}^2 - 2u p_r^2}= \sqrt{1 + (1 - 2u) p_r^2 + \frac{{\mathbf L}^2}{\mu^2 R^2}}\,.
\ee
In the limit $G\to 0$, i.e. $u\to 0$ (keeping, however, fixed the centrifugal energy term ${\mathbf L}^2/(\mu^2 R^2)$ which does not
contain a factor $G$), we get $W=w_p$, so that 
\be\label{gssnu0G0}
\lim_{G\to0} \lim_{\nu\to0} g_{S_*}=\frac{1}{1+w_p}  +\frac{1}{w_p}=\frac{ \left( 1+2w_p \right) }{ w_p (1+w_p) }\,.
\ee
It is easy to check that the $\nu \to 0$ limits  of our 1PM results \eqref{gsfin}, \eqref{gssfin}, agree with the
corresponding expressions \eqref{gsnu0}, \eqref{gssnu0G0}.

As a second check on our results, let us compare them to the fractionally $O(1/c^4)$ accurate PN expansions of 
$g_S $ and $g_{S_*} $ \cite{Nagar:2011fx,Barausse:2011ys}. To make this comparison, we need, however, to use an
anti-DJS spin gauge. Let us recall, that the results of Refs. \cite{Damour:2008qf,Nagar:2011fx,Barausse:2011ys} involve
some arbitrary gauge parameters. [As already mentioned, this arbitrariness is linked to introducing a
time-dependent rotation of the local frame in which the spins are measured \cite{Damour:2008qf}.] 
From  Eqs. (29), (30) in \cite{Nagar:2011fx}, we have the structure
\begin{align}
g_S^{\rm PN}     = 2 & + \dfrac{1}{c^2}g_S^{{\rm NLO}}(a) + \dfrac{1}{c^4}g_S^{\rm NNLO}(a;\alpha,\beta,\gamma),\\
g_{S_*}^{\rm PN}  = \dfrac{3}{2} &+ \dfrac{1}{c^2}g_S^{{\rm NLO}}(b) + \dfrac{1}{c^4}g_{S^*}^{{\rm NNLO}}(b;\delta,\zeta,\eta), 
\end{align}
where the gauge parameters entering $g_S $ are called $a$ (which enters at $O(1/c^2)$) and $\alpha, \beta, \gamma$ 
(entering at $O(1/c^4)$), while the corresponding gauge parameters entering $g_{S_*} $
are called $b$ (which enters at $O(1/c^2)$) and $\delta, \zeta, \eta$ (entering at $O(1/c^4)$). The explicit form of these PN
expansions read
\begin{widetext}
\begin{align}
 \label{eq:gSeff_full}
  g_S^{\rm PN}=2&+\dfrac{1}{c^2}\Bigg[\left(\dfrac{3}{8}\nu+a\right)\p^2-\left(\dfrac{9}{2}\nu+3 a\right)(\n\cdot\p)^2\Bigg)
                - u(\nu+a)\Bigg]\nonumber\\
               &+\dfrac{1}{c^4}\Bigg[- u^2\left(9\nu + \dfrac{3}{2}\nu^2 + a + \alpha \right)
                   \nonumber\\
                  &+u\left[(\n\cdot\p)^2 \left(\dfrac{35}{4}\nu - \dfrac{3}{16}\nu^2 + 6a - 4\alpha-3\beta-2\gamma\right)
                   +\p^2\left(-\dfrac{17}{4}\nu + \dfrac{11}{8}\nu^2 - \dfrac{3a}{2}+\alpha-\gamma\right)\right]\nonumber\\
                   &+ \left(\dfrac{9}{4}\nu - \dfrac{39}{16}\nu^2 + \dfrac{3a}{2} +3\beta-3\gamma \right)   \p^2 (\n\cdot \p)^2
                    + \left(\dfrac{135}{16}\nu^2-5\beta\right)                                              (\n\cdot \p)^4   \nonumber \\
                   &+\left(-\dfrac{5}{8}\nu-\dfrac{a}{2} + \gamma\right)                                    \p^4             
                    \Bigg],\\  
 g_{S_*}^{\rm PN}  =\dfrac{3}{2}&+\dfrac{1}{c^2}\Bigg[\left(-\dfrac{5}{8}+\dfrac{1}{2}\nu+b\right)\p^2 
                 -\left(\dfrac{15}{4}\nu+3b\right)(\n\cdot\p)^2- u\left(\dfrac{1}{2}+\dfrac{5}{4}\nu+b\right)\Bigg]\nonumber\\ 
    &+\dfrac{1}{c^4}\Bigg[- u^2\left(\dfrac{1}{2} + \dfrac{55}{8}\nu + \dfrac{13}{8}\nu^2 + b+ \delta\right)
                     \nonumber\\
                   &+ u\Bigg[(\n\cdot \p)^2\left(\dfrac{5}{4}+\dfrac{109}{8}\nu + \dfrac{3}{4}\nu^2+6 b-4\delta-3\zeta-2\eta\right)
                    +\p^2\left(\dfrac{1}{4}-\dfrac{59}{16}\nu + \dfrac{3}{2}\nu^2 - \dfrac{3b}{2}+ \delta-\eta\right)\Bigg]\nonumber\\
                   & +\left(\dfrac{57}{16}\nu - \dfrac{21}{8}\nu^2 + \dfrac{3b}{2} + 3\zeta - 3\eta\right)     \p^2(\n\cdot\p)^2
                    +\left(\dfrac{15}{2}\nu^2                       - 5\zeta\right)                            (\n\cdot \p)^4\nonumber\\
                   &+\left(\dfrac{7}{16}-\dfrac{11}{16}\nu-\dfrac{\nu^2}{16}-\dfrac{b}{2} + \eta \right) \p^4
                    \Bigg].
\label{eq:gSstareff_full}
\end{align}
It is easily checked that we can move to an anti-DJS gauge (i.e. eliminate the dependence of  $g_S$ and $g_{S*}$  on $p_r$ to keep
a dependence only on $\p^2$ and $u$), by the following choice of gauge parameters:
\begin{eqnarray}
&&a = -\frac{3}{2}\nu\,,\qquad\alpha = -\frac{1}{16}\nu-\frac{7}{4}\nu^2\,,\qquad \beta = \frac{27}{16}\nu^2\,,\qquad \gamma = \frac{7}{8}\nu^2,\nonumber\\
&&b = -\frac{5}{4}\nu\,,\qquad\eta = \frac{9}{16}\nu+\frac{5}{8}\nu^2\,,\qquad \delta = \frac{5}{16}+\frac{5}{4}\nu-\frac{5}{4}\nu^2\,,\qquad \zeta = \frac{3}{2}\nu^2\,.
\end{eqnarray}
This yields the following results,
\begin{eqnarray}
g_S^{\rm PN} &=& 2+\frac{1}{c^2} \left(-\frac{9}{8}\nu \p^2+\frac12 u\nu \right)\nonumber\\
&+& \frac{1}{c^4}\left[-u^2 \left(\frac{119}{16}\nu-\frac14 \nu^2\right)+u \p^2 \left(-\frac{33}{16}\nu-\frac{5}{4}\nu^2\right)+\left(\frac18\nu+\frac{7}{8}\nu^2\right)\p^4\right]\nonumber\\
g_{S_*}^{\rm PN}  &=&  \frac32 +\frac{1}{c^2}\left[\left(-\frac{5}{8}-\frac{3}{4}\nu\right) \p^2-\frac12 u\right]\nonumber\\
&+& \frac{1}{c^4}\left[-u^2 \left(\frac{13}{16}+\frac{55}{8}\nu+\frac{3}{8}\nu^2\right)+u \p^2 \left(\frac{9}{16}-\frac{9}{8}\nu-\frac{3}{8}\nu^2\right)
+\left(\frac{7}{16}+\frac12 \nu+\frac{9}{16}\nu^2\right) \p^4\right]\,.
\end{eqnarray}
\end{widetext}
The PM re-expansion of these (PN-expanded) expressions amounts to ordering them in powers of $u=GM/R$, i.e.
\begin{eqnarray}
g_S &=&  g_S^{\rm 1PM}(\p^2)+u\,  g_S^{\rm 2PM}(\p^2) + u^2\,  g_S^{\rm 3PM}(\p^2) ,\nonumber\\
g_{S_*} &=&   g_{S_*}^{\rm 1PM}(\p^2)+u\, g_{S_*}^{\rm 2PM}(\p^2)+ u^2\,  g_{S_*}^{\rm 3PM}(\p^2),
\end{eqnarray}
with
\begin{eqnarray} \label{gs1PMPN}
g_S^{\rm 1PM}(\p^2)&=&  2-\frac{9}{8c^2}\nu \p^2+\frac{1}{c^4}\left(\frac{1}{8}\nu+\frac{7}{8}\nu^2\right)\p^4\nonumber\\ 
&+& O(\p^6),\nonumber\\
g_{S_*}^{\rm 1PM}(\p^2)&=&  \frac32 +\frac{1}{c^2}\left(-\frac{5}{8}-\frac{3}{4}\nu\right) \p^2\nonumber\\
& +&\frac{1}{c^4} \left(\frac{7}{16}+\frac12 \nu+\frac{9}{16}\nu^2\right) \p^4\nonumber\\
&+& O(\p^6),
\end{eqnarray}
\bea
g_S^{\rm 2PM}(\p^2)&=& \frac{1}{2c^2}\nu+\frac{1}{c^4} \p^2 \left(-\frac{33}{16}\nu-\frac{5}{4}\nu^2  \right) + O(\p^4), \nonumber\\
g_{S_*}^{\rm 2PM}(\p^2)&=&  -\frac{1}{2 c^2}+\frac{1}{c^4} \p^2 \left(\frac{9}{16}-\frac{9}{8}\nu-\frac{3}{8}\nu^2\right) \nonumber\\
& +& O(\p^4)\,,
\eea
and
\begin{eqnarray}
g_S^{\rm 3PM}({\mathbf p}^2)&=&  \frac{1}{c^4} \left(-\frac{119}{16}\nu+\frac14 \nu^2\right) +O({\mathbf p}^2),\nonumber\\
g_{S*}^{\rm 3PM}({\mathbf p}^2)&=&  \frac{1}{c^4} \left( -\frac{13}{16}-\frac{55}{8}\nu-\frac38 \nu^2 \right)\nonumber\\
& +& O({\mathbf p}^2)\,.
\end{eqnarray}
It is straightforward to check that the PN expansion (i.e. the expansion in powers of $\p^2$) of our exact 1PM results \eqref{gsfin}, \eqref{gssfin}, agree with the PN results \eqref{gs1PMPN} (within their $O(\p^6)$ accuracy). 

Let us finally briefly mention  the knowledge of the gravitational self-force (SF) expansion 
(i.e. the expansion in powers of $\nu$) of $g_{S} $ and $g_{S_*} $.
Several recent papers (notably Refs. \cite{Bini:2014ica,Bini:2015xua,Kavanagh:2017wot}) have investigated the first term ($\propto \nu$) in the SF expansion of $g_{S} $ and $g_{S_*} $. However, these SF studies have been limited to the case of circular, or slightly-eccentric bound orbits. Because of this fact,
one cannot directly compare our close-to-hyperbolic results to such SF results. Let us, however, mention an interesting aspect of
our 1PM results \eqref{gsfin}, \eqref{gssfin}, namely their behavior at ultrarelativistic energies. First, let us remark that, in the extreme-mass-ratio
limit $\nu \to 0$, we have the following large-energy behavior when the relative Lorentz $\gamma$ factor $w=\sqrt{ 1  +{\mathbf p}^2 }$
tends to $\infty$:
\be
\left[g_S\right]_{\nu=0}=2 \,; \,\left[g_{S_*}\right]_{\nu=0} \sim \frac{2}{w} ; \, {\rm as} \,\,  w \to \infty\,.
\ee
By contrast, in the comparable-mass case, i.e. when $\nu \neq 0$, we have the large-energy behaviors
\be
g_S \sim \frac{2}{\nu \, w} \,; \, g_{S_*} \sim \frac{\sqrt{2}}{\sqrt{\nu} \, w^{3/2}} ; \, {\rm as} \,\,  w \to \infty\,.
\ee
In words, this means that comparable-mass effects drastically change the large-energy behavior of $g_S$ and $g_{S_*}$.
When $\nu \neq 0$, both gyrogravitomagnetic ratios tend to zero at large energies, and $g_{S_*}$ tend to zero faster than before.
The large-energy behavior of $g_S$ and $g_{S_*}$ (both for $\nu=0$ and for $\nu\neq0$) are illustrated in Fig. 3.
\begin{figure}
\includegraphics[scale=0.35]{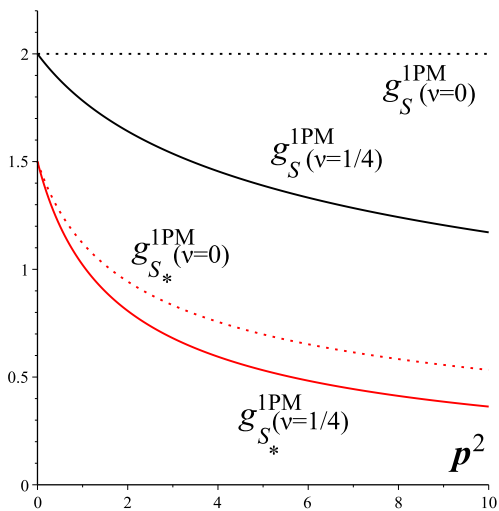}
\caption{\label{fig:3}  Our 1PM results, Eqs. \eqref{gsfin},  \eqref{gssfin}, for the EOB spin-orbit couplings $g_S^{\rm 1PM}({\mathbf p}^2,\nu)$ and  $g_{S*}^{\rm 1PM}({\mathbf p}^2,\nu)$ are plotted as functions of $\p^2$ for the two limiting values of the symmetric mass-ratio $\nu=0$ and $\nu=\frac14$.}
\end{figure}

This illustrates the {\it singular} character of the SF expansion at large energies. Actually, we see on Eqs. \eqref{gsfin}, \eqref{gssfin}
that the extra $\nu$-related factors embodying these faster large-energy decays consists of a factor $\sim 1/ (1+ 2 \nu (w-1))$ in $g_S$,
and a factor $\sim 1/ \sqrt{1+ 2 \nu (w-1)}$ in $g_{S_*}$. Performing the formal SF expansion of $g_S$ and $g_{S_*}$
would mean expanding these factors in powers of $\nu$, according to, e.g.
\be \label{sfexp}
\frac{1}{\sqrt{1+ 2 \nu (w-1)}} = 1 - \nu (w-1) + O(\nu^2) \,.
\ee
We see that, while the exact PM-type lhs goes to zero as $w \to +\infty$, its first-order SF-expanded version on the rhs goes towards $- \infty$
 as $w \to +\infty$. Such a singular behavior of SF-expanded quantities was first observed in Ref. \cite{Akcay:2012ea}
for orbital effects, and was also found for spin-orbit effects in Refs. \cite{Bini:2014ica,Bini:2015xua}.
However, in the SF context, the large-energy regime is reached when considering orbits near the light-ring of a black hole.
As the latter near-light-ring limit mixes large-kinetic-energy effects with strong-field effects, one cannot directly 
translate our specific large-energy PM effects into a corresponding light-ring behavior. Still, we think that our PM approach
illuminates the issue by showing the explicit appearance of factors involving powers of  $1/ \sqrt{1+ 2 \nu (w-1)}$,
which generate singular SF terms of the type shown on the rhs of Eq. \eqref{sfexp}.

\section{Conclusions}

A new gauge-invariant approach to the description of spin-orbit coupling in binary systems has been introduced.
It is based on the new, related, concepts of ``scattering holonomy" (integrated connection along an entire
hyperbolic-motion worldline), and ``spin holonomy" (action of the scattering holonomy on the spatial spin three-vector).
We have formulated our approach in the approximation where we neglected spin-curvature effects (corresponding,
in the Hamiltonian approach, to nonlinear-in-spin effects), but it can be generalized (by modifying the evolution law
of the spins) to the inclusion of spin-curvature effects. Compared to the approach (suggested in \cite{Damour:2016gwp})
consisting in including  spin-orbit effects in the computation of the scattering angle, the approach presented here
has the significant advantage that we can derive information on spin-orbit effects from a calculation where we actually 
neglect spin-effects ! [In that respect, this is akin to the method used in Refs. \cite{Damour:2007nc,Damour:2008qf}.]

We have applied here our method  to the explicit computation of the spin-orbit couplings at the first order
in the post-Minkowskian expansion (first order in $G$ and all orders in $v/c$). Using then an extension
of the EOB/real dynamics dictionary, we have transcribed our results into the computation of the
two gyrogravitomagnetic ratios $g_S$ and $g_{S_*}$, see Eqs. \eqref{gsfin}, \eqref{gssfin}.
Our results are compatible with the previous knowledge on these coupling coefficients, but 
extend our knowledge in the direction of arbitrarily-high momenta. In particular, it has been found that,
for comparable-mass binary systems ($\nu \neq 0$), $g_S^{\rm 1PM}({\mathbf p}^2,\nu)$ and  $g_{S*}^{\rm 1PM}({\mathbf p}^2,\nu)$
tend to zero in the ultrarelativistic limit (see Fig. 3). Our work provides new insights on the singular
nature of the self-force expansion. 
We leave to future work the exploitation of our results in the currently physically more urgent
case of black-hole coalescences (ellipticlike motions, instead of the hyperboliclike ones considered here). 
Our all-orders-in-$(v/c)$ results can suggest new ways of resumming the spin-orbit couplings.
Let us note that our finding that $g_S$ and $g_{S_*}$ decay at large kinetic energies,
resonate with the finding that the fitting of EOB theory to numerical relativity data indicates 
a significant decay  of $g_S$ and $g_{S_*}$ during the strong-field coalescence of binary black holes
(see, e.g., the calibration of the spin-orbit parameters $d_{\rm SO}$ in \cite{Taracchini:2012ig,Taracchini:2013rva},
and $c_3$ in Refs. \cite{Damour:2014sva,Nagar:2015xqa}).
It will be interesting to extend our results to the second post-Minkowskian level ($O(G^2)$) to complete
our information about the regime where both  kinetic and the binding energies become large.

\section*{Acknowledgments}
We are grateful to Ulysse Schildge for fruitful exchanges during many years.
D.B. thanks the International Center for  
Relativistic Astrophysics Network (ICRANet) and the  
Italian Istituto Nazionale di Fisica Nucleare (INFN) for  
partial support and the Institut des Hautes Etudes 
Scientifiques (IHES) for warm hospitality at various stages  
during the development of the present project.

\end{document}